# Semantic Modeling with Foundries

Robert B. Allen and Yoonhwan Kim
Yonsei University, Seoul, South Korea
rba@boballen.info, lumiere_au@naver.com

**Abstract:** We analyze challenges for the development of the Human Activities and Infrastructures Foundry. We explore a rich semantic modeling approach to describe two Korean ceramic water droppers used to mix ink for calligraphy, how they were produced and the reasons for their differing aesthetic. Our modeling supports schema and allows for transitions of Entities based on the relationships to other Entities with which they are associated. We explore the similarity of our approach to object-oriented analysis and modeling.

## 1    INTRODUCTION

Direct representation digital libraries are based on highly structured knowledgebases such as the Basic Formal Ontology (BFO). BFO is a data representation framework that is derived from Aristotelian realist principles and is widely used in biomedicine. BFO [8] distinguishes between Universals and Particulars as well as among Occurrents and Independent and Dependent Continuants. Ontologies based on BFO are being collected into Foundries. The Open Biomedical Ontology (OBO) Foundry [20] is well established, and there is an emerging Industry Ontologies Foundry (IOF).[1] In [6], we proposed the Human Activities and Infrastructures Foundry (HAIF). Here, we consider in detail some of the issues for the implementation of that Foundry. Further, in previous work (e.g., [3]), we have proposed that BFO has similarities to object-oriented modeling, such as in the central role of inheritance (cf., semantic data model) and the close coupling of Occurrents and Independent Continuants (cf., methods and classes). Here we explore additions to the foundry ontologies to support modeling.

## 2    SEMANTIC MODEL LAYER

Ontologies identify the types of Entities that make up the Reality. The goal of modeling is to use Entities to create representations of a dynamic Reality. The emphasis of both approaches on describing the world with Entities suggests the similarities between them should be explored. For instance, BFO proposes that every Process must be associated with at least one Independent Continuant (e.g., an Object). This is remarkably similar to the pairing of objects with methods, which is the goal of encapsulation in object-oriented approaches. Thus, we explore how BFO version 2.0 (BFO2) may be interpreted and extended to make it compatible with object-oriented analysis, design, and programming, especially in the context of specifying the HAIF [6]. Cumulatively, these methods form an extension of BFO2 that we have described as a Model Layer.

### 2.1    Objects

BFO2 specifies Objects as Material Entities and, therefore, as Independent Continuants. An application ontology may or may not provide links between Material Entities and Physical Properties and Locations, but we believe it should be clear that Object "has" them. If the Object has Parts, then we believe the Function of those Parts should be included. In short, we propose the use of structured schemas to define Thick Objects.[2] These structures would be similar to Java Classes and FrameNet Frames. Indeed, the active vocabulary used for an application could be compiled much as a computer program is compiled.

### 2.2    Object Aggregates

Object Aggregates ([8] p93) are a collection of relatively co-equal spatially-distributed Material Entities. For instance, a symphony orchestra is considered an Object Aggregate because its members, the players, are all Material Entities.[3] If we find even one member of the Aggregate then we can instantiate the slots for the entire Aggregate. The key to the Object Aggregate is not its members but the network of their relationships. A full account should characterize complex interactions among the components (e.g., the players). Though, this can be at progressive levels of detail. In cases such as kinship, recursion may be used at part of the definition.

It is informative to compare the BFO2 distinction between Objects and Object Aggregates with the operational definition of Parts in the Unified Modeling Language (UML). In UML Parts are said to have a Composition relationship to a Class if they are destroyed when the Class is destroyed. Otherwise, the Parts are said simply to be Contained by the Class.[4] Another related distinction, between Open and Closed Systems, comes from General Systems Theory (cf., [3]). In all these cases, the definitions of the Processes and Transitionals also affect the details of these distinctions.

### 2.3    Qualities

In BFO2 Qualities are Dependent Continuants, which Inhere_In Independent Continuants. Qualities such as Color and Mass associated with Material Entities are "determinables" ([8] p97), and are required Qualities. Specific values for determinable Qualities are determinants (e.g., red color). A model should distinguish between determinables and determinants. Indeed, determinants could form their own "quality ontology" (e.g., the ontology for colors of a traffic light would be "green," "yellow," and red").

---

[1] https://www.youtube.com/watch?v=y0TeTfoFdSA
[2] We might extend this to all Independent Continuants.
[3] Non-material systems, such as a system of laws or a computer model, would appear not to be included under the BFO2 definitions.
[4] This is consistent with the distinction made by BFO2. If an Object such as a clock is destroyed, its parts generally are destroyed, but if a symphony orchestra is destroyed presumably its members may survive. We also argue that a Systems view would provide a useful alternative to the primarily structuralist approach of anatomy [10].

Some other Qualities are Relational Qualities ([8] p97). These allow relationships between Independent Continuants to be first-class objects. For instance, between two people there may be a "married to" relationship.

## 2.4 Realizables

While Determinant Qualities are permanently associated with an Object, other Dependent Continuants may only sometimes be realized. In BFO2, these are called Realizables and they are subdivided as Roles, Dispositions, and Functions.

Roles are associated with the relationships among the Objects that comprise Relational Qualities and Object Aggregates. For instance, [11] develops a BFO analysis of a town in Nebraska that focuses on the activities of the person who has the Role of Superintendent of Schools. For many formal roles, Organization Theory and Coordination Theory [12] may be usefully applied.

Dispositions are a type of Realizable that suggest the potential for a physical change in a Material Entity. "… a Disposition is a realizable entity … such that if it ceases to exist, then its bearer is physically changed" ([8] p101). For example, "a car windshield has a sure-fire disposition to break if it is struck with a sledgehammer moving at 100 feet per second…" ([8] p6).[5]

Functionality has long been recognized to be important for description. Functional analysis, functional requirements, and functional models are integral to Business Process Modeling (BPM) and object-oriented models. In BFO2, Functions are Dispositions that are designed for a specific purpose. For instance, tools can be defined by their Functions. Thus, the Function of a pencil is to write. Furthermore, Functions are often linked both laterally (a piston rod is connected to the crankshaft) and vertically (the function of writing). Following a functionality to higher levels of abstraction, a Function is often said to help satisfy human Needs. A Need may be said to be a derived Entity required for the development, operation and maintenance of a System, as in Communication serves the Needs of a social system (Section 3.3).

An important point is that Roles and Functions are often applied to rich contexts (e.g., Roles are associated with Object Aggregates). A rigorous application of Realizables should make those associations explicit.

## 2.5 Processes, State Changes, and Entity Transitions (Transitionals)

BFO2 defines Processes as Occurrents. Processes may be characterized as a continuous activity by an Independent Continuant. Indeed, a full description of a Process must include the Independent Continuants to which it may be applied. Processes do not themselves allow State Changes; instead, BFO2 includes Process Boundaries ([8] p123) which may help to account for State Changes. However, Process Boundaries are said to be instantaneous and therefore have zero dimensions in both space and time. We do not see how something that has zero dimensions in both space and time can be considered real. Therefore, we have proposed that State Changes be a derived model-level construct which we designate Transitionals [4]. Such Transitionals provide a bridge between ontologies and executable statements in programming languages.

We implement Transitions as atomic operations[6] that create and/or delete Relationships of Objects to other Entities (cf., [2, 4, 6, 9]).. For instance, for a traffic light to change color, the light Color relationship is moved from Red to Green. In an object-oriented interpretation, Relationships between Entities suggest that there may be "message passing" between them.[7]

## 2.6 Transition Chains

BFO2 allows for the possibility of Process Chains where, for instance, the stages of development from an embryo is seen as a Process, without specifying the details of the transitions between stages. There are several ways that Transitions may be connected. A Transition Chain may be identified as simply a disjoint sequence, a mechanism, a procedure, or as a workflow. The differences among them are sometimes small and the terms are often used interchangeably but we can make useful distinctions.[8] Mechanisms [14] and Procedures can be either Universals or Particulars. When they are Universals, they can be like programs and program methods. Thus, they may include flow control such as conditionals and loops. The Universals can then be instantiated and run as Particulars.

Just as we proposed to use Thick Objects for Independent Continuants (Section 2.1), we could have Thick Chains which would provide a structured description of the entire Chain to the extent it has been specified. Moreover, we should be able to compare different Mechanisms and determine, for instance, whether they are functionally equivalent.

## 2.7 Microworld

In addition to describing the interactions within parts of a given Thick Object, we can also consider interactions between different Thick Objects. In BFO2, such interactions are associated with Object Aggregates and involve changes to relationships such as Relational Qualities or Roles. To support concurrent interactions, it is helpful to define a virtual space, a Microworld in which those interactions occur.[9]

---

[5] However, we might, instead, consider the Disposition for glass to break as the Quality of impact strength. A distinction might be made that impact strength would normally be measured by some action on the Object (an impact). Moreover, it might be possible to calculate impact strength directly from physical properties.
[6] These are similar to productions for OPS5 and contracts in Eiffel in programming languages.
[7] At least in some cases, these complex interactions could be considered as a type of Role. See the discussion of the Release Frame in [2].

[8] As the term suggests, a Mechanism would be a self-contained machine (e.g., a clock mechanism) while Procedures and Workflows would have some immediate external control or intervention (typically by a human being). It is also worth distinguishing between planned and completed Mechanisms and Procedures.

## 3 KOREAN WATER DROPPERS

We study two museum objects to identify some of the differences between descriptions of Universals and Particulars, and to explore the possibility of using the techniques described here to supplement traditional metadata. Specifically, we considered two Korean ceramic water droppers (shown below). A water dropper holds water for the calligrapher to drip onto an ink stone. Then, an ink stick is rubbed into the water to blend ink to the desired consistency. On the left is a celadon water dropper from the Goryeo Dynasty (918-1392 AD) in the shape of a duck.[10] On the right is an elegant and austere, ring-shaped water dropper from the Joseon Dynasty (1392-1897).[11]

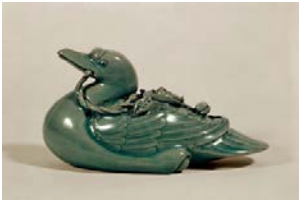 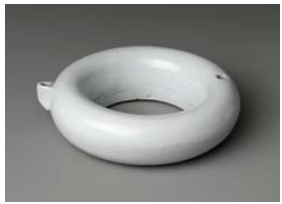

### 3.1 Material Entities in a Social Context

Everyday human activates and infrastructures interactions are complex and we need multiple perspectives for describing them. [6] proposed three levels for describing social or cultural objects.[12]

The first level is purely physical. For pottery, this level would describe the Materials and Procedures for production and use, as well as physical attributes of the finished object. For instance, the description of the celadon water dropper might include the relevance of oxidation/reduction reactions for glazes and kiln construction. We could also model the Procedure[13] for making the pottery. Throwing it changes its shape (a Quality) and firing it changes its moisture content. In addition, other procedures related to the water droppers such as mixing the ink when using the water dropper, ink stone, dry ink stick, and brush, could be modeled. As suggested in Section 2.5, such procedures could be defined in general terms, specialized, and possibly instantiated.

### 3.2 Routine Social Structures and Interaction

A second level models the social structure of a community on its own terms. This might include model elements for religion, governance, kinship, and education. For the water droppers, we might describe the social structure of the village in which they were produced. The area around Gangjin County along the southern coast of Korea was the center of celadon production. It could be viewed as a Spatio-Temporal region and as a political Entity. The pottery was produced by workers and craftsmen and distributed through merchants. We also need to connect social roles to tasks such as the Roles associated with kiln procedures,[14] and could describe the social structure including the scholars (Yangban) who do the calligraphy.

At yet another level, we could describe the Procedures of the museum(s) which collected and display the objects. The timeline for these events can be a "unified temporal map".

### 3.3 Higher-Level Social Processes

A third level models social meta-structures. This might include models to explain the reasons for the social structures described in Section 3.2. Such analysis may invoke sociological theories. Functionality and Systems are central to some of these theories.[15] Of course, social structures change. For instance, the difference in style of the two water droppers may reflect major cultural differences between the Dynasties during they were made. One explanation for the adoption of a more austere style in the Joseon Dynasty is as a rejection of what was seen as the excesses of Goroyeo aristocrats and the acceptance of NeoConfucian principles. More deeply, there was a complex of interacting factors including economics and politics. As with any history, we cannot know all of the facts but structured descriptions of the changing social structures over time should match natural language and help illuminate the history.

## 4 FOUNDRY DESIGN

To populate Foundries, individual ontologies need to be developed at scale. For instance, repetitive semantic structure [16] could be implemented programmatically. In biology, because of the wide variety of species biological ontologies, as well as biological research itself, often focus on model organisms such as *Drosophila* or *e. coli*. We propose the similar development of model communities,[17] although the challenges of variability and exceptions in the types of communities may equal those among types of organisms [19].

The ontologies in Foundries should be modular so that they can be effectively managed and applied as needed. Certainly, different perspectives and levels of analysis should be well represented. However, there is a danger of losing continuity across components (cf., [13]) and, worse, uncontrolled propagation of component ontologies. As suggested in Section 3.1, applications may require terms from several different Foundries and, ultimately, the several Foundries should be unified.

---

[9] In programming terms, the Microworld would be a blackboard and the interaction of Objects in the Object Aggregate could be implemented with multiple dispatch.

[10] Korean National Treasure #74, bit.ly/2CfJMOD

[11] www.metmuseum.org/art/collection/search/72620

[12] These three levels are somewhat related to Panofsky's distinction between icon, iconography, and iconology [17].

[13] A Procedure for the production of celadon is posted on the Gangjin County history webpage: http://152.99.169.51:8080/en/contents.do? key=1931

[14] [18] contains descriptions of life in a 12th century celadon pottery village.

[15] [3] briefly considered Parsons's Structure-Functionalism that considered the Needs of the social system. Parsons is sometimes criticized for adopting a rigid view of homeostasis in the social system. In this high-level overview, we do not take a position on that and would accept a weaker form of homeostasis. We might also consider Malinowski's Functionalism [16] which focuses on the Needs of individuals.

[16] Consider: baker, baking, bakery; brewer, brewing, brewery [2].

[17] The analogy to Model Organisms is useful but not without challenges as Model Organisms seem to be both an ideal and exemplar. Perhaps this is because in different contexts, a species can be either a Universal or a Particular.

The BFO ontologies are a type of classification systems for entities. The different perspectives provided by the foundry ontologies and across some of the Dependent Continuants create a form of faceting. Moreover, traditional measures of exhaustivity and specificity for library classification and indexing systems are directly applicable.

While logic and functional programming may seem a natural approach for BFO [9], we believe that an object-oriented approach is also useful. To do that, we introduce a model layer that extends the BFO framework.[18] Just as the object-oriented modeling approach is applied to HAIF ontologies, it could be applied to industry ontologies and biology ontologies. However, we recognize that several paradigms may contribute and we are currently implementing semantic modeling in Python.[19] Python is a flexible language with functional, imperative, and object-oriented modes.

## 5 DISCOURSE THREADS

In addition to semantics of the HAIF, we also need to develop ways of interacting with the contents of the foundry. That includes both policies for building the foundry and for using the foundry to describe Particular communities. That is, we need to support claims with evidence and argumentation. Typically, evidence involves using some Material Entity or Information Artifact whose authenticity can be validated together with a relevance model for how it supports the claim. Those assertions can then be used in explanations and narrative. In the case of history, many details are unknown and probably can never be known. In those cases, historians may revert to applying what they consider to be the best fitting Mechanism or Procedure.

Some sense of causation is needed, at least implicitly, for different types of discourse. Explanations which use scientific models (e.g., chemical reactions and Newtonian mechanics) depend on the validity of those models and their codification as accepted knowledge [5]. In other cases, causation is a matter of contentious argument. While we accept INUS formulation [15] about the role of necessary and sufficient conditions in explaining causation, the specification of those conditions may be subjective and depends on the available vocabularies and models.

## 6 CONCLUSION

We have explored issues in the development of HAIF and examined some of the issues associated with rich descriptions of cultural objects. We have proposed that ontologies be more closely coordinated with modeling languages. From the perspective of modeling languages, such an approach has the potential to extend their focus beyond business process modeling to social and cultural processes. For instance, we can consider applying techniques traditionally associated with BPM, such as UML and systems analysis, in the development of community, societal and cultural models.

While there remain open issues, direct representation may enable a new generation of digital libraries and of interacting with complex scholarly materials. Similar techniques may be applied to a broad range of structured descriptions. For instance, they could be used to develop a rich-semantic version of DBpedia. They could also be used to develop interactive narrative timelines (e.g., [1]) and even interactive historical games and dramas.

---

[18] This may be said to form an eXtended Formal Ontology (XFO).

[19] Allen, R.B., and Jones, T., Semantic Modeling Programming Environment, in preparation.